# A non-spherical core in the explosion of supernova SN 2004dj

Douglas C. Leonard[1,3], Alexei V. Filippenko[2], Mohan Ganeshalingam[2], Franklin J. D. Serduke[2], Weidong Li[2], Brandon J. Swift[2], Avishay Gal-Yam[1], Ryan J. Foley[2], Derek B. Fox[1,3], Sung Park[2], Jennifer L. Hoffman[2], Diane S. Wong[2]

[1]*Astronomy Department, MS 105-24, California Institute of Technology, Pasadena, California 91125, USA*

[2]*Department of Astronomy, University of California, Berkeley, California 94720-3411, USA*

[3] Present addresses: Department of Astronomy, San Diego State University, San Diego, California 92182, USA (D.C.L.); Department of Astronomy and Astrophysics, Pennsylvania State University, 525 Davey Lab, University Park, PA 16802, USA (D.B.F.)

**An important and perhaps critical clue to the mechanism driving the explosion of massive stars as supernovae is provided by the accumulating evidence for asymmetry in the explosion. Indirect evidence comes from high pulsar velocities[1], associations of supernovae with long-soft $\gamma$-ray bursts[2,3], and asymmetries in late-time emission-line profiles[4]. Spectropolarimetry provides a direct probe of young supernova geometry, with higher polarization generally indicating a greater departure from spherical symmetry[5,6]. Large polarizations have been measured for 'stripped-envelope' (that is, type Ic; ref. 7) supernovae, which confirms their non-spherical morphology[8,9]; but the explosions of massive stars with intact hydrogen envelopes[7,10] (type II-P supernovae) have shown only weak polarizations at the early times observed[11,12]. Here we report multi-epoch spectropolarimetry of a classic type II-P supernova that reveals the abrupt appearance of significant polarization when the inner core is first exposed in the thinning ejecta (~ 90 days after explosion). We infer a departure from spherical symmetry of at least 30 per cent for the inner ejecta. Combined with earlier results, this suggests that a strongly non-spherical explosion may be a generic feature of core-collapse supernovae of all types, where the asphericity in type II-P supernovae is cloaked at early times by the massive, opaque, hydrogen envelope.**

SN 2004dj was discovered[13] on 31 July 2004 (UT) in the nearby spiral galaxy NGC 2403, and was quickly identified[14] as a core-collapse event with spectral features similar to those of the type II-P supernova SN 1999em (ref. 10) about three weeks after explosion. At an estimated distance of 3.13 Mpc (ref. 15), it is the closest normal type II-P supernova observed to date. We present spectropolarimetry spanning the first nine months of its development in Table 1 and in Figs 1 and 2.

The light curve of SN 2004dj (Fig. 2) exhibits the classic features of a type II-P supernova: an enduring period of nearly constant optical luminosity followed by a precipitous drop in brightness and a smooth decline on the radioactive tail. The first three epochs of spectropolarimetry, obtained during the middle and end of the plateau phase, find the supernova to be unpolarized (Fig. 1), indicating a substantially spherical geometry for the electron-scattering atmosphere while the photosphere is receding back through the hydrogen envelope.

A global asphericity of the electron-scattering atmosphere will manifest itself through two main spectropolarimetric signatures. First, there is significant polarization in spectral regions not dominated by line opacity, resulting from a non-balancing of the electric vectors of the scattered continuum photons[5,8]. For type II-P supernovae, this includes the broad spectral region 6800 – 8200 Å (ref. 20). Very little polarization should be seen in spectral regions shortward of wavelength $\lambda \approx 5500$ Å owing to line-blanketing by Fe; other spectral regions with strong individual emission lines (for example, H$\alpha$ and Ca II near-infrared triplet) should also be relatively unpolarized[16]. Second, polarization increases are anticipated in the troughs of strong P-Cygni absorption lines[11,21]. This results from selective blocking of more forward-scattered and, hence, less polarized, light in the trough regions. A further expectation resulting from a global asphericity, as opposed to other potential supernova polarization mechanisms (see ref. 17 for an extensive discussion), is that no rotation of the polarization angle (PA) should be seen across these absorption-line line features.

Once the ionized hydrogen envelope has recombined, the optical brightness of a type II-P supernova drops dramatically as the recombination front rapidly recedes through the helium core[20]. Previous studies[12] have noted the confluence of factors making the drop-off of the plateau a propitious period in which to search for evidence of explosion asphericity, as the innermost regions are exposed at a point when sufficient optical depth to electron scattering in the inner core still exists to polarize the light and, hence, reveal the geometry. Our fourth spectropolarimetric observation captures this moment in a type II-P supernova (to our knowledge, for the first time). Taken during the steepest part of the descent off the plateau (Fig. 2), on day 91 after explosion, the data show a dramatic change from those obtained just 19 days earlier: the polarization level has increased to nearly 0.6% in the continuum-dominated region 6800 - 8200 Å (Fig. 1), as well as in the troughs of some of the stronger P-Cygni line features, including most prominently that due to Na I D. The PAs in the P-Cygni troughs agree with the continuum PA, with a measured value of $\theta = 32° \pm 3°$ in the Na I D absorption compared with $\theta = 28° \pm 2°$ for the continuum region. The observed spectropolarimetric behavior of SN 2004dj matches expectations resulting from a strong departure from spherical symmetry for the electron-scattering atmosphere in the inner regions of the ejecta (expansion velocities less than ~2,500 km s$^{-1}$, or about 1/10 the radius of the supernova assuming a maximum ejecta velocity of ~25,000 km s$^{-1}$).

As SN 2004dj moves into the nebular phase and the atmosphere continues to expand, the optical depth to electron scattering, and thus the polarization, is expected to fall. Specifically, if the day 91 observation occurred at or beyond the time when the single-scattering limit is reached in the supernova's atmosphere (a single-scattering atmosphere produces maximum supernova polarization[6]), then the polarization should subsequently decrease in proportion to the (angle-averaged) electron-scattering optical depth, as this determines the percentage of scattered light. For this simple scenario, the polarization would be expected to decline with time $t$ as $1/t^2$, owing to geometrical dilution caused by homologous expansion (that is, the same number of scattering electrons spread over a cross-sectional area that grows as $t^2$). This prediction is borne out rather well by the data (Fig. 2), although there is some evidence suggestive of a more complex situation. Specifically, the PA changes by 30° between days 91 and 128, after which it remains constant for the next three epochs (Table 1). This result has a high degree of statistical significance, and indicates that the scattering environment's morphology might be more complicated than that of a single, fixed geometry; we speculate on possible causes of this PA rotation below.

The link between long-soft $\gamma$-ray bursts (GRBs) and a subset of type Ic supernovae -- specifically those with spectral lines indicating very high expansion velocities at early times[4] -- is now secure[2,3], and successful models for GRBs require a jet-like explosion from energetics considerations. While it is not clear that the same mechanism that generates the GRB is also responsible for exploding the star, the substantial early-time polarization that has been measured for some of these events[9], and more typical type Ic supernovae as well[8], supports the contention of a non-spherical explosion for 'stripped-envelope' supernovae. The most thoroughly studied non-spherical explosion model is that of bipolar jets[22], which distribute radioactive $^{56}$Ni and other synthesized heavy elements (such as Fe) preferentially along the polar axes; lighter elements produced by the progenitor star during its evolution (for example, O) are distributed near the equatorial plane in a disc-like structure[23]. This model, in fact, has recently been successfully used to explain the emergence of double-peaked oxygen line profiles in spectra of SN 2003jd, a type Ic supernova with high early-time expansion velocities, obtained a year after explosion[4].

For type II-P supernovae, the bipolar jet model predicts the radioactive $^{56}$Ni to be confined to the inner regions, where it would ionize the surrounding material, produce a non-spherical electron-scattering photosphere when exposed[24], and plausibly produce polarization similar to that observed in SN 2004dj. In addition, it has been argued[25] that a non-spherical distribution of $^{56}$Ni can produce asymmetries in the H$\alpha$ emission-line profile of type II-P supernovae during the nebular phase. Such asymmetries have, in fact, been recently identified in late-time spectra of SN 2004dj, and are interpreted within the context of the bipolar jet model by ref. 19, which concludes that they are best explained by a bipolar distribution of $^{56}$Ni with the major axis oriented $30° \pm 10°$ to our line-of-sight. If we adopt this viewing orientation, then our measured polarization peak of 0.56% places a minimum departure from spherical symmetry of ~30% (axis ratio 10:7) for the inner regions of SN 2004dj from the axially symmetric, electron-scattering models of ref. 6. Naturally, if the electron-scattering atmosphere on day 91 has not yet reached the single-scattering limit, or is already below it, then the actual degree of asphericity is greater than our derived value.

The lack of polarization during the plateau fits comfortably within the bipolar jet model as well, as hydrodynamical simulations demonstrate that the shock propagates laterally as it penetrates the inner regions[23], and ultimately becomes completely spherical as it proceeds through the massive hydrogen envelope[26]. Because the shock propagation (and, hence, $^{56}$Ni production) in the bipolar jet model is not purely radial in the inner regions, a rapidly but smoothly changing PA might be expected during the brief period (SN 2004dj falling off the optical plateau) in which the $^{56}$Ni is being uncovered. This could explain the PA rotation observed for SN 2004dj between days 91 and 128, although opacity effects may also play a role[27]. Additional examples, preferably with fine temporal sampling during the critical phase at the end of the plateau, are needed to fully test these conjectures.

Analysis of pre-explosion images places the initial main-sequence mass of SN 2004dj's progenitor at 12 solar masses[18], in accord with masses inferred by similar studies of other type II-P supernovae (ref. 28 and references therein). Theory predicts that single stars that explode as type Ic supernovae must have initial main-sequence masses greater than ~25 solar masses to undergo the extraordinary wind-driven mass loss required to remove their entire H/He envelopes prior to explosion[29], and observations of Galactic Wolf-Rayet stars support this conclusion[30]. The prompt appearance of substantial polarization in type Ic supernovae supports a non-spherical explosion model for these very massive stars. The discovery of a non-spherical core in SN 2004dj suggests that a similar explosion mechanism is at work in type II-P supernovae, and may, in fact, be inherent to the core-collapse process, regardless of progenitor mass.

We thank Daniel Kasen for useful discussions. Support for this research was provided by the National Science Foundation, NASA, and the Sylvia and Jim Katzman Foundation. A.V.F. is grateful for a Miller Research Professorship at U.C. Berkeley, during which part of this work was completed.

**Correspondence** and requests for materials should be addressed to D.C.L. (e-mail: leonard@sciences.sdsu.edu; ph: 619-594-2215).

**Table 1 | Spectropolarimetric Observations of SN 2004dj**

| Epoch (days) | Total Exposure (s) | p(%) ±1$\sigma$ | $\theta$(°) ±1$\sigma$ |
|---|---|---|---|
| 39 | 600 | $0.079 \pm 151$ | ND |
| 58 | 9,200 | $0.000 \pm 041$ | ND |
| 72 | 2,360 | $-0.006 \pm 041$ | ND |
| 91 | 5,200 | $0.558 \pm 042$ | $28 \pm 2$ |
| 128 | 3,600 | $0.401 \pm 054$ | $178 \pm 4$ |
| 157 | 19,200 | $0.215 \pm 042$ | $173 \pm 6$ |
| 187 | 14,400 | $0.141 \pm 052$ | $178 \pm 11$ |
| 242 | 9,200 | $0.071 \pm 047$ | $171 \pm 20$ |
| 271 | 5,600 | $0.052 \pm 058$ | ND |

Epochs are with respect to the estimated date of explosion, 2004 July 14 (ref. 14). Continuum polarization level, $p$, and polarization position angle in the plane of the sky, $\theta$, represent measurements made on the ISP-subtracted data (see Fig. 1 legend) over the range 6800 – 8200 Å; in an effort to avoid statistical biases that exist at low polarization levels, the reported polarization is the median of the rotated Stokes parameter[11] over this interval. The PA is determined as $\theta = 1/2 \times \tan^{-1}(u_{med}/q_{med})$, where $u_{med}$ and $q_{med}$ are the median values of the Stokes parameters over this interval. The day 39 data were acquired using the double spectrograph with polarimeter on the 5-m Hale Telescope at Palomar Observatory. All other data were obtained with the Kast double spectrograph with polarimeter at the Cassegrain focus of the Shane 3-m telescope at Lick Observatory. The uncertainties in the Stokes parameters, which were then used to derive the uncertainties in $p$ and $\theta$, are derived as the quadrature sum of the Poisson error and a term reflecting our estimate of the random uncertainty in the overall, measured level of the Stokes parameters. For the Lick data, we set this term to be 0.04%, which is the average 1σ (s.d.) spread in the measured Stokes parameters of null standards observed on multiple nights (the same nights as the SN 2004dj data were obtained). For the day 39 data obtained at Palomar, we set this term equal to 0.15%, to account for the likely existence of some instrumental polarization in the data: while null standards at Lick were always found to be null to within 0.1%, the null standards observed at Palomar on this night possessed measured polarizations of up to 0.25%, in a way that could not be consistently modelled or removed from the data. Note that values for $\theta$ are ill-defined for the day 39, 58, 72, and 271 data owing to the low polarization. On day 271, a final 2,367 seconds of observations (a third complete set of four waveplate positions) were acquired but not included when forming the combined Stokes' parameters for the night's observations, as they were taken under deteriorating weather conditions and with poor observing technique (object accidentally placed near the edge of the slit), and are discrepant with the earlier data at the 4σ (s.d.) level. ND, not determined.

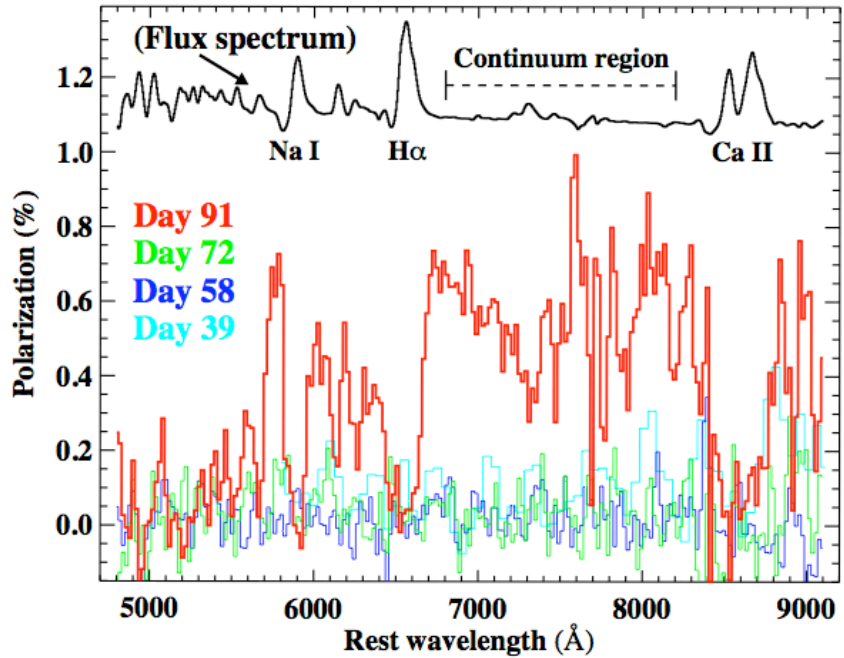

**Figure 1. Intrinsic polarization of SN 2004dj during the photospheric phase with day since explosion indicated.** The flux spectrum is from day 91, with prominent line features labelled. The recession velocity of 131 km s$^{-1}$ (NASA Extragalactic Database) has been removed from all data. The polarization data have been corrected for an interstellar polarization (ISP) contribution, which was derived by examining (1) the observed polarization in spectral regions dominated by the strongest emission lines (that is, Na I D, H$\alpha$ and the Ca II near-infrared triplet), which are believed to be intrinsically unpolarized[16], and (2) the observed polarization at late times (that is, the nebular phase), when the electron-scattering optical depth of the atmosphere has dropped well below unity and the observed polarization at all wavelengths is largely due to ISP (see ref. 17 for a complete discussion). These complementary approaches yield virtually identical results (to within 0.1%) for the ISP estimation. For our data set, the final epoch is from day 271, at which point the ejecta are probably not fully nebular and some small intrinsic polarization (<0.1%) may still exist. Examining the temporal evolution in the Stokes q-u plane, we find the late-time polarization to be steadily decreasing down to the level indicated by the earliest measurements made during the plateau phase, suggesting that these early epochs possess little to no intrinsic polarization (this is also supported by the lack of any line features at these times). Thus, to derive the intrinsic polarization at all epochs, we fitted a low-order spline to the Stokes *q* and *u* spectra of the data from day 58 (note that the data from day 39 have significantly lower signal-to-noise ratio than those obtained at

the other plateau epochs), and then subtracted the result from the Stokes parameters obtained on all other epochs. This resulted in the removal of $p_{ISP} \approx 0.29\%, \theta_{ISP} \approx 20°$ from the observed data.

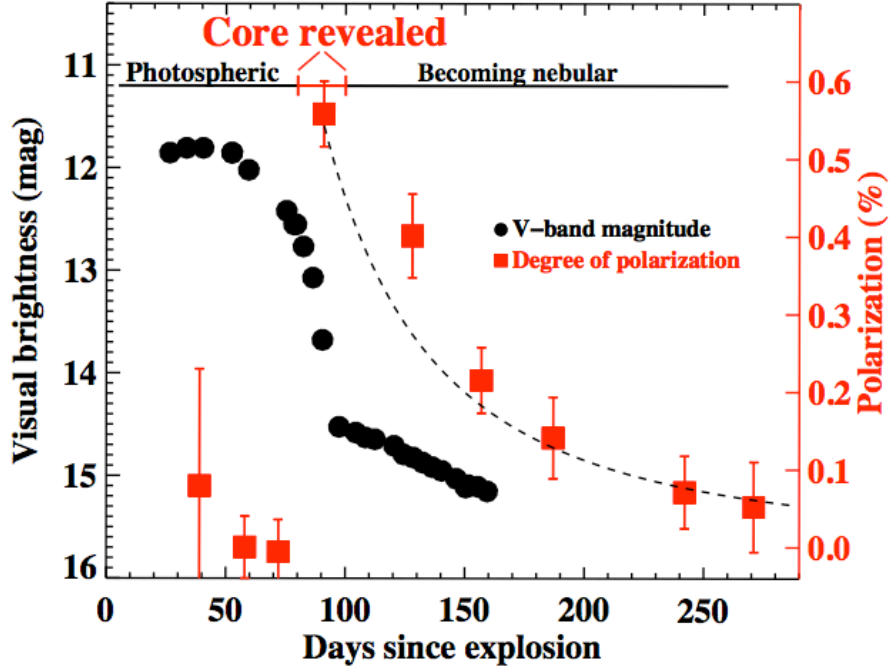

**Figure 2. Light curve and continuum polarization of SN 2004dj.** Polarization measures are from Table 1; error bars are 1σ (s.d.) statistical. Photometry is from data obtained with the 30-inch (0.8-m) Katzman Automatic Imaging Telescope (KAIT) at Lick Observatory and the 60-inch (1.5-m) telescope at Palomar Observatory. The dashed line represents the expected decline in polarization ($p = (t_0/t)^2 \times p_0$, where $t_0 = 91$ days and $p_0 = 0.558\%$ from Table 1) during the transition to the nebular phase due to the effects of diminishing electron scattering in optically thin, expanding ejecta. Note that the age of SN 2004dj at discovery and, hence, the plateau duration, is not well constrained by direct observation, as NGC 2403 had just emerged from solar conjunction and the most recent reported pre-explosion image was taken over six months earlier[18]. Our adopted explosion date of 2004 July 14 results from the spectral analysis of ref. 14, but we note that the light-curve modeling of ref. 19 yields an explosion date 31 days earlier. If the earlier date were adopted, the estimated plateau duration would increase from ~70 to ~100 days.